\documentstyle[10pt,aaspp4,tighten,psfig]{article}

\lefthead{Graham {\it et al.}}
\righthead{NGST IFTS}

\def\wig#1{\mathrel{\hbox{\hbox to 0pt{%
  \lower.5ex\hbox{$\sim$}\hss}\raise.4ex\hbox{$#1$}}}}

\begin{document}

\title{THE PERFORMANCE AND SCIENTIFIC RATIONALE FOR
AN INFRARED IMAGING FOURIER TRANSFORM SPECTROGRAPH
ON A LARGE SPACE TELESCOPE}

\author{James R. Graham }
\affil{
       	Department of Astronomy\\
	601 Campbell Hall\\
	University of California\\
	Berkeley, CA 94720\\
       (jrg@astro.berkeley.edu)
	}

\author{Mark Abrams}
\affil{ITT Aerospace/Communications Division\\
	1919 West Cook Road, P.O. Box 3700\\
	Fort Wayne, IN 46801\\
	(mcabrams@itt.com)
	}

\author{C. Bennett}
\affil{Lawrence Livermore National Laboratory\\
	MS L-043\\
	P.O. Box 808\\
	Livermore, CA 94550\\
	(bennett2@llnl.gov)
	}

\author{J. Carr}
\affil{Naval Research Laboratory\\
	Code 7217\\
	4555 Overlook Ave.\\
	Washington, DC 20375\\
	(carr@mriga.nrl.navy.mil)
	}

\author{K. Cook}
\affil{Lawrence Livermore National Laboratory\\
	MS L-041\\
	P.0. Box 808\\
	Livermore, CA 94550\\
	(kcook@llnl.gov)
	}

\author{A. Dey}
\affil{Dept. of Physics \& Astronomy\\
       The Johns Hopkins University\\
       3400 N. Charles St.\\
       Baltimore, MD 21218\\
       (dey@pha.jhu.edu)
	}

\author{J. Najita}
\affil{Space Telescope Science Institute\\
	3700 San Martin Drive\\
	Baltimore, MD 21218\\
	(najita@stsci.edu)
	}

\author{E. Wishnow}
\affil{ Lawrence Livermore National Laboratory
\& Space Sciences Laboratory\\
University of California Berkeley\\
MS L-401\\
P.O. Box 808\\
Livermore, CA 94550\\
(wishnow@llnl.gov)
}

%JOURNAL WILL TAKE CARE OF THE FOLLOWING
%\received{}
%\accepted{}
%\journalid{}{}
%\articleid{}{}

\slugcomment{Submitted to: 
Publications of the Astronomical Society of the Pacific}

\vfill

\begin{abstract}

We describe a concept for an imaging spectrograph for a large orbiting
observatory such as NASA's proposed Next Generation Space Telescope
(NGST) based on an imaging Fourier transform spectrograph (IFTS).  An
IFTS has several important advantages which make it an ideal
instrument to pursue the scientific objectives of NGST. We review the
operation of an IFTS and make a quantitative evaluation of the
signal-to-noise performance of such an instrument in the context of
NGST. We consider the relationship between pixel size, spectral
resolution, and diameter of the beamsplitter for imaging and
non-imaging Fourier transform spectrographs and give the condition
required to maintain spectral modulation efficiency over the entire
field of view.  We give examples of scientific programs that could be
performed with this facility.

\end{abstract}

\keywords{ instrumentation: interferometers and spectrographs ---
techniques: interferometric and spectroscopic}

%\twocolumn

\section{INTRODUCTION}

The Association of Universities for Research in Astronomy (AURA), with
NASA support, recently appointed a committee to ``study possible
missions and programs for UV-Optical-IR astronomy in space for the
first decades of the twenty-first century.''  The report urged the
development of a general-purpose, near-infrared observatory equipped
with a passively cooled primary mirror ($T \le 70$~K) with a minimum
diameter of 4 meters (Dressler 1996). To enhance its performance, the
report recommended that the observatory be placed as far from the
Earth-Moon system as possible to reduce stray light and to maintain
the telescope's relatively low temperature. With such a facility,
it should be possible to learn in detail how galaxies formed, measure the
large-scale curvature of space-time by measuring distant
standard-candles, trace the chemical evolution of galaxies, and study
nearby stars and star-forming regions for signs of planetary
systems. A detailed discussion of the next generation space
telescope (NGST) and its scientific potential given by Stockman
(1997).

For NGST to attain these scientific objectives, it must have an
instrument which is designed to execute panchromatic observations over
the critical 1--15 $\mu$m wavelength range of the faintest detectable
objects.  With nJy sensitivity levels attainable at near-infrared
(1--5 $\mu$m) (NIR) and mid-infrared (5--15$\mu$m) (MIR) wavelengths,
NGST will be able to study the well-calibrated rest-frame optical
diagnostics in distant ($z=3-10$) galaxies, thus probing for the first
time, their stellar content, star-formation history and nuclear
activity.  At the longer wavelengths, NGST can investigate these
properties in $z=3-5$ galaxies using diagnostics that are unaffected
by dust extinction and reddening, and also study the dust properties
directly.

At the flux limits characteristic of NGST, the confusion limit is
likely to be approached, with virtually every pixel having significant
information (e.g., by extrapolation from counts in the Hubble Deep
Field (Williams et al. 1996)).  As a result, one of the best ways to
maximize the scientific output from NGST is to provide a wide-field
imaging spectrograph that is efficient in this limit.

An imaging Fourier transform spectrometer (IFTS) provides these
capabilities in a low-cost, high throughput, compact design.  It
provides the only efficient means of conducting {\it unbiased}
spectroscopic surveys of the high-$z$ Universe, i.e., without object
preselection (e.g., using broad band colors) and without the
restrictions imposed by 
spectrometer slit geometry and placement.  An IFTS also
allows spectroscopy over a wide bandpass, affords flexibility in
choice of resolution, is easy to calibrate, and is ideal for
wide-field spectroscopic surveys.  Bennett et al. (1993) and Bennett,
Carter, \& Fields (1995) describe the operating principles of imaging
Fourier transform spectrographs and compare their performance with
alternative imaging spectrometers.  A comprehensive review of the
application of interferometers and the techniques of Fourier
spectroscopy to astrophysical problems is given by Ridgway \& Brault
(1984), and a recent summary of the field, including a description of
an astronomical IFTS is given by Maillard (1995).

Spaceborne Fourier transform spectrometers have been responsible for
spectacular results in the fields of planetary exploration and
cosmology.  Infrared FT spectrometers developed at the Goddard Space
Flight Center (GSFC) flew on board the Mariner 9 mission to Mars, and
were carried to the outer planets by the Voyager spacecraft (Hanel et
al. 1992).  The instruments provided superb data revealing, for the
first time, the composition of the atmospheres of the giant gaseous
planets (e.g. Jupiter, Hanel et al. 1979).  The Composite Infrared
Spectrometer (CIRS), currently traveling to Saturn onboard the
Cassini spacecraft, is another instrument developed at GSFC.  CIRS
is the first step towards an imaging FTS as it has a linear array of
detectors, rather than a single element detector, in order to map the
temperature and composition of the atmospheres of Saturn and Titan as
a function of altitude during limb soundings.  (Kunde et al. 1996).

The definitive measurement of the spectrum of the cosmic microwave
background radiation (CMBR) was one of the most dramatic experimental
measurements of this decade (Mather et al. 1990; Gush, Halpern \&
Wishnow 1990).  The FIRAS instrument onboard the NASA satellite COBE
which first performed this measurement, and the COBRA rocket
experiment conducted by the University of British Columbia which
confirmed it a few months later, were both liquid-helium cooled,
differential Fourier transform spectrometers.  These instruments used
a dual-input, dual-output configuration where one input viewed the sky
and the other viewed a blackbody calibrator (Mather, Fixen \& Shafer
1993; Gush \& Halpern 1992).  Absolute photometric measurements were
obtained by reference to the blackbody calibrator, and the CMBR was
observed to have an undistorted Planck spectrum corresponding to a
temperature of $2.728\pm0.004$ K (Fixen et al. 1996).  The IFTS
proposed here can be thought of as an extension of these experiments
where focal plane detector arrays yield simultaneous imaging and
spectral information.

In the next decade, missions such as WIRE, AXAF, and SIRTF will expand
astrophysical horizons, possibly unveiling entirely new populations of
objects.  An IFTS offers the flexibility (e.g., spectral resolution)
that may prove essential in investigating the nature of these sources.
Due to its flexibility and its ability to provide simultaneous imaging
and spectroscopy of every object in the field of view (FOV), an IFTS
is a {\it necessary} instrument for the NGST mission.

\section {IFTS CONCEPT}
\label{concept}

\begin{figure}[thb]
\centerline{\psfig{figure=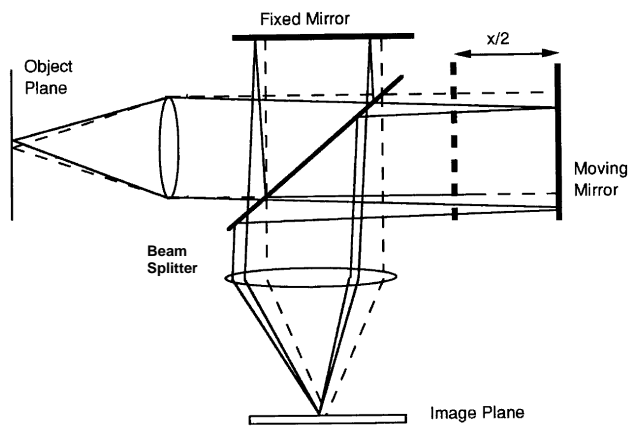,width=3.3in,angle=-90}}
\caption{\small A sketch of the optics of a simple single-beam
imaging Fourier transform spectrograph consisting of
a collimating lens, a beam splitter, two mirrors (one movable),
and a camera lens.
The optical path difference is $x$.}
\label{fts}
\end{figure}

An IFTS (Fig. \ref{fts}) is axis-symmetric, and the optical path
difference (OPD) is the same for all the points of the image with the
same angle of incidence from the axis of the interferometer. Hence,
the FOV is circular. On the object side, an entrance collimator
illuminates the interferometer with parallel light. The interfering
beams are collected by the output camera, creating a stigmatic
relation between the object and image planes. By placing a detector
array in the output focal plane the entrance field is imaged on the
array and each pixel works as a single detector matched to a point on
the sky.

Retrieving spectral information involves recording the interferogram
generated by the source imaged onto the focal plane array (FPA). The
OPD is scanned in discrete steps since FPAs are integrating detectors.
Scanning in this way generates a data cube of two-dimensional
interferograms.  The signal from the same pixel in each frame forms an
independent interferogram.  These interferograms are Fourier
transformed individually yielding a spectral data cube composed of the
same spatial elements as the image.

\subsection{A Perfect Match to NGST Science}
\label{perfect}

\begin{figure}[thb]
\centerline{ \psfig{figure=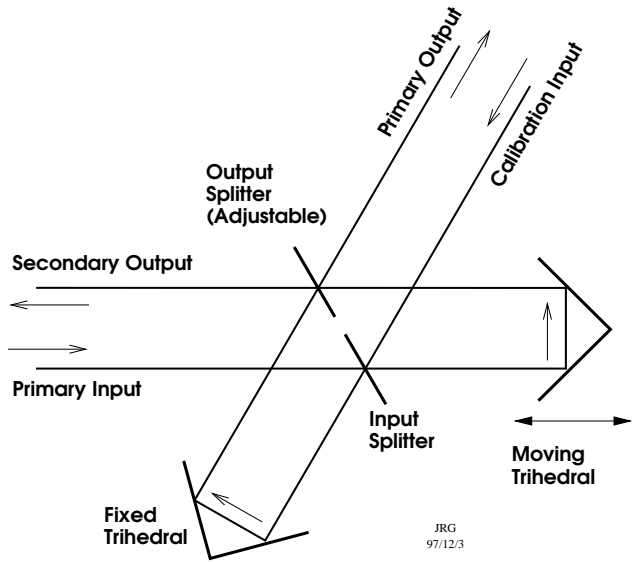,width=3.3in,angle=-90}}
\caption{\small
Schematic optical layout of a 60$^\circ$
dual-input, dual-output Michelson interferometer.
}
\label{optical-layout}
\end{figure}

The features of an IFTS which make it the instrument of choice for
NGST are efficiency, flexibility, and compactness. The most compelling
reason for choosing an IFTS is that in the dual port design (see
Fig. \ref{optical-layout}) virtually every photon collected by the
telescope is directed towards the focal plane for detection. Other
solutions are inefficient, inflexible, and wasteful of mass, power,
and volume. Cameras equipped with filters admit only a restricted
bandpass at low spectral resolution. To compete with the spectral
multiplex advantage of an IFTS, a camera system needs multiple
dichroics and FPAs. The additional mass and thermal load is a severe
penalty. Classical dispersive spectrographs have slit losses, grating
inefficiencies due to light lost in unwanted orders, and limited free
spectral range (the same is true for a Fabry-Perot).  {\em An IFTS
acquires full bandpass imaging simultaneously with higher spectral
resolution data.} Therefore, a high SNR broad-band image always
accompanies full spectral sampling of the FOV with no penalty in
integration time.  An IFTS is a true imaging spectrograph and measures
a spectrum for every pixel in the FOV.  It is not necessary to choose
which regions in the image are most deserving of spectroscopic
analysis.  Overheads are eliminated because no additional observing
time is needed for imaging prior to object selection, and there is no
delay in positioning slit masks, fibers, or image slicing
micro-mirrors. Thus, an IFTS will produce a rich scientific legacy with
tremendous potential for serendipity.

Table \ref{capabilities} details the capabilities of a putative IFTS
suitable for NGST.  We use the instrument described by this table to
illustrate the potential of an IFTS.  Two points in Table
\ref{capabilities} must be stressed: 1) An IFTS is spectrally
multiplexed, therefore all spectral channels are obtained
simultaneously within the stated integration time. 2) The free
spectral range of an IFTS is limited only by the band-pass filter and
the detector response. Consequently, the usual definition of
resolution, $R = \lambda/\delta \lambda$, is of limited use. It is
conventional to scan the OPD of an IFTS in equal steps so that the
resolution is constant in wavenumber, $k$. Thus, we use $M$ to
denote the number of spectral channels. For example, in the NIR with a
1-5 $\mu$m band-pass, $M=5$ means that $\delta k = (k_{max} -
k_{min})/M = 1600$~cm$^{-1}$, and a scan yields 5 bands centered 1.1,
1.3, 1.7, 2.3, \& 3.6 $\mu$m.

%\onecolumn

\begin{deluxetable}{lll}
\tablewidth{0pt}
\tablecaption{Capabilities of a Putative NGST IFTS}
\tablehead{
\colhead{} & \colhead{NIR Channel} &\colhead{MIR Channel}}
\startdata
Design			& Dual-port & Dual-port \\
Bandpass		& 1-5 $\mu$m & 5-15 $\mu$m \\
Resolution  		& 1 cm$^{-1}$	& 1 cm$^{-1}$ \\ 
FOV 			& $200''$	& $100''$ \\
Pixel size 		& $0.''05$	& $0.''1$ \\
Array format 		& 4k$\times$4k & 1k$\times$1k \\
Detector		& InSb 		& HgCdTe \\
Throughput		& $> 0.5$	& $> 0.5$ \\	
Sensitivity\tablenotemark{a} \\		
~$M=1$		& 200 pJy & 13 nJy \\
~$M=5$			& 1 nJy   & 65 nJy  \\
~$M=100$		& 35 nJy  & 1.3 $\mu$Jy \\
\enddata
\tablenotetext{a}{SNR = 10 for a $10^5$~s 
integration over the entire  spectral 
band for a point source.
$M$ is the number of simultaneous
spectral channels in the band-pass ---  see \S \ref{perfect}.
Note that all
spectral channels are obtained simultaneously.
The spectrum is assumed to be flat in $F_\nu$ and the
SNR is quoted at 2 $\mu$m for the NIR channel, and
at 10 $\mu$m for the MIR channel.}
\label{capabilities}
\end{deluxetable}

%\twocolumn

\begin{figure}[thb]
\centerline{\psfig{figure=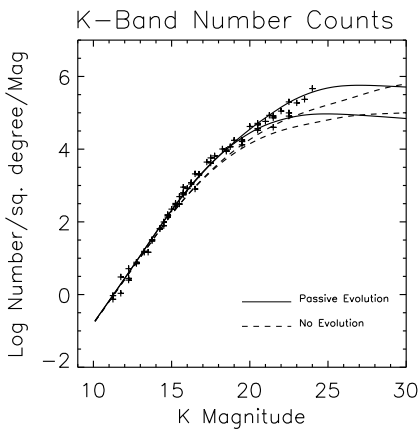,width=3.4in,angle=90}}
\caption{\small $K$-band number counts (Djorgovski et al. 1995; Gardner et
al. 1993, 1996; Glazebrook et al. 1994; Huang et al. 1997; Mobasher et
al. 1986; Moustakis et al. 1997; McLeod et al. 1995; \& Metcalfe et
al. 1996) together with models of the luminosity function modeled
using the formalism of Gardner (1998), which has been used to
extrapolate the number counts into the NGST domain. The solid lines
include the effects of passive evolution, while the dashed lines
include only K-corrections. The upper line in each case is for $q_0 =
0.1$, and the lower lines are for $q_0 = 0.5$.  Current number counts
imply at least 3500 objects per $3.'3$ NGST field, while the
extrapolations shown here suggest as many as 11,000 to $K=29.5$.  }
\label{number-counts}
\end{figure}

The throughput of an IFTS with ideal optics is only limited by the
efficiency of the beam splitter.  In a dual-input, dual-output port
design no light is wasted and the throughput approaches 100\%.  An
IFTS has no loss of light or spatial information because there is no
slit, hence an IFTS is perfectly adapted to doing multi-object
spectroscopy in crowded or confusion limited fields.  A IFTS uses
every photon whereas traditional cameras and spectrographs throw away
photons (either spectrally with a filter or spatially with a slit), so
at a very fundamental level an IFTS is superior.  On blaze, a good
grating is 80\% efficient, but averaged over the free spectral range
this drops to about 65\%.  An IFTS is not optimized for single-object
spectroscopy because the broad-band photon shot noise is associated
with every frame in the interferogram. Hence, for a single object a
slit spectrograph is $\eta_g \eta_s M$ times faster than an IFTS of
the same resolution in background limited operation, where $\eta_g$ is
the grating efficiency averaged over the blaze function, and $\eta_s$
is the slit loss, where typically the product $\eta_g \eta_s \approx
0.3$.  This disadvantage is more than compensated for by the
spatial-multiplexing capability of an IFTS. A typical deep background
limited exposure of an IFTS will reach $K=29.5$, SNR=10 and will
contain at least 3500 and possibly, depending on cosmology, up to
11,000, objects per field (see Fig. \ref{number-counts}).  A grating
spectrograph with a fiber feed or multi-slit capability can perhaps
record spectra for only a few percent of these objects at a time,
requiring hundreds of pointings to make an unbiased survey of a single
field, as opposed to the single IFTS imaging-cum-spectroscopic
observation.

An IFTS is tolerant of detector noise because it always operates under
photon limited conditions due to the broad spectral bandpass
transmitted to the FPA. This is illustrated in Table
\ref{readoutexamples} which shows a break-down of the noise sources in
the NIR and MIR channels corresponding to the performance listed in
Table \ref{capabilities}.  Table \ref{readoutexamples} also shows that
the read-out rates required to avoid saturation are modest
($1-10$~mHz), since typical well depths for NIR InSb or HgCdTe arrays
are a few $10^5$ e$^-$ and $10^7$ e$^-$ for MIR Si:As arrays.

%\onecolumn

\begin{deluxetable}{llll}
\tablewidth{0pt}
\tablecaption{Signal and Noise Budget\tablenotemark{a}}
\tablehead{
\colhead{} & 
\colhead{} &
\colhead{NIR Channel} 
&\colhead{MIR Channel}}
\startdata
       &       & $F$ = 1 nJy & $F$ = 65 nJy \\
       &       & $t$ = 1000 s & $t$ = 100 s \\
       &       & $\Delta\lambda = 1-5\mu$m & $\Delta\lambda = 5-15\mu$m \\
\hline
Signal \\
       & Source     &  610 & 2709   \\
       & Background\tablenotemark{b} & 3724 & 463669 \\
       &            &      &        \\
Total Signal  &            & 4334 & 466378 \\
\\
Noise\\
           & Signal Shot Noise  & 24.7 & 52.0 \\
           & Background Shot Noise & 60.8 & 680.9 \\
	   & Dark Shot Noise   & 5.5  & 10.0 \\
	   & Read Noise        & 5    & 5 \\
Total Noise &           & 66.0 & 683.0 \\
\enddata
\label{readoutexamples}
\tablenotetext{a}{In electrons}
\tablenotetext{b} {Background includes zodiacal foreground
and thermal emission from the telescope as calculated as described in
\S \ref{snrcal}.}
\end{deluxetable}

%\twocolumn

Similarly, orders of magnitude higher thermal emission from the
instrument, or thermal radiation leaks from outside the instrument
bay, can be tolerated compared to the case for dispersive
spectrometers or fixed filter cameras.  As a pragmatic demonstration
of this principle, the IFTS instruments LIFTIRS and HIRIS are
routinely operated with ambient temperature optics in the 8-14 $\mu$m
band (Bennett et al. 1995), whereas dispersive spectrometers, like
SEBASS (Bennett, {\it private communication}), operating in the same
spectral region must have the slit and all following optics cooled far
below ambient temperatures.  The reason is that in a dispersive
spectrometer the thermal emission of all the elements and optics
downstream of the slit reach the detector at full spectral range
determined by the bandpass limiting element at or near the coldstop,
whereas only the narrow spectral range corresponding to the width of a
spectral channel for the signals of interest reach the detector
pixels. For the IFTS, both the signals of interest and the thermal
emission are seen over the full spectral range determined by the
bandpass limiting filter, and thus it is only necessary that the
thermal emission of the optical elements along the optic axis
integrated over the bandpass of interest, either 1-5 $\mu$m or 5-15
$\mu$m, be somewhat less than that of the integral of the zodiacal
foreground, telescope emission, and source signal level integrated
over the same broad spectral region.

An IFTS is potentially immune to cosmic ray hits because the
``energy'' of a single upset pixel in one OPD frame appears as a
sinusoidal signal divided among all bins in the spectral transform of
the interferogram for that pixel.  We can ignore cosmic ray hits only
if the counts generated are at or below our noise level. A minimum
ionizing cosmic ray proton ($E \simeq 1$ GeV) has ionization losses of
$dE/dx \simeq 400$ eV $\mu$m$^{-1}$ in Si.  Assuming that 3.6 eV is
required to produce an electron-hole pair, a cosmic ray will
yield at least a few thousand events, since typical pixels have
sensitive layers that are tens of $\mu$m thick. 
We would obtain similar number for a hybrid device, i.e.,
InSb or HgCdTe on a Si multiplexer.
If a cosmic ray hit
produces a significant signal in a certain number of pixels, those
pixels must be ``repaired'' by interpolating the interferogram between
the previous few ``good'' frames, and the following few ``good''
frames which are not contaminated by cosmic ray hits. The same sort of
processing would be needed for any other system as well, be it an
imager or a spectrometer. Comparison with the noise sources listed in
Table \ref{readoutexamples} indicates that cosmic ray hits will have
to be repaired in the NIR channel, while the MIR channel will be more
tolerant.

A dual port design (Fig. \ref{optical-layout}) delivers the
complementary symmetric and antisymmetric interferograms.  In this
dual-input dual-output design, the field of the complementary input
(labeled ``Calibration Input'' in Fig. \ref{optical-layout}) is also
imaged and superimposed on each image of the ``Primary Input''.  This
property is often used to cancel the sky emission.  In operation, when
observing the sky in the primary input, the secondary input would be
fed with a cold blackbody load, having negligible radiance.  The final
interferogram is constructed from the difference 
between the two outputs (which is therefore
also immune to common mode electrical noise) while the normalized
ratio reveals systematic variation due to detector drifts.

The wavelength scale and the instrumental line shape (a $sinc$
function if there is no apodizing) are precisely determined and are
independent of wavelength. Absolute wavelength calibration is done by
counting fringes of an optical single-mode laser.  Compared to a
dispersive system the broad-band operation of an IFTS means that there
are $M$ times more photons for flat fielding and determining
signal-dependent gain (linearity). Hence, high signal-to-noise
calibration images can be acquired faster or with lower power internal
sources.

\subsection{Pixel Size, Spectral Resolution, and Field of View}
\label{pixelsize}

Spatial multiplexing renders the performance of an IFTS equal to that
of an ideal multi-slit spectrograph (Bennett 1995). Hence, even if we
ignore slit losses and blaze inefficiency the other advantages of an
IFTS are overwhelming.  The spectral resolution can be varied
arbitrarily from the coarsest case of a small number of bands up to a
spectral resolution limit determined only by the maximum OPD
characteristic of the instrument. The proposed instrument has a
maximum OPD of 1 cm and hence can operate over a range of resolutions
from full band up to $M$=8000 in the NIR

The spectral resolution limit, $R = k/\delta k$, of a Michelson
interferometer is

\begin{equation} 
R = 8(d/ \phi D)^2, 
\end{equation}

\noindent
where $\phi$ is the angular diameter of the FOV, $d$ is the diameter
of the beam splitter, and $D$ the telescope primary mirror diameter
(e.g., Jacquinot 1954; Maillard 1995). Classically, $\phi$ refers to
the entire field,
but in the case of an IFTS, $\phi$ is the FOV of an on-axis
pixel.

Although convenient if a single fringe fills the FPA, just as
with imaging Fabry-Perots, there is no reason why each pixel should
record the same apparent wavenumber.  Fringes crowd together with
increasing field angle.  Therefore, the need to maintain modulation
efficiency over the entire field of view requires that the 
spatial separation of
the fringes at the edge of the FPA, for a given
retardance, is significantly greater than the pixel spacing. 

If $x$ is the OPD for a normally incident beam with wavenumber $k$,
and $\theta$ is the field angle of off-axis rays at the beam splitter,
then the path difference at $\theta$ is $x_\theta = x cos\theta$ and
the apparent wavenumber of this beam is 

\begin{equation}
k_\theta = k/cos\theta.
\end{equation}

\noindent
The angles $\theta$ and $\phi$ are related by the angular
magnification, $D/d$. If $\delta \theta$ is the angular width, also at
the beam splitter, corresponding to a single pixel, the spectral
resolution limit for off-axis points can be found by differentiating
Eq (2),

\begin{equation}
1/R_\theta = \delta k_\theta/k_\theta = tan\theta \delta \theta, 
\end{equation}

\begin{figure}[thb]
\centerline{\psfig{figure=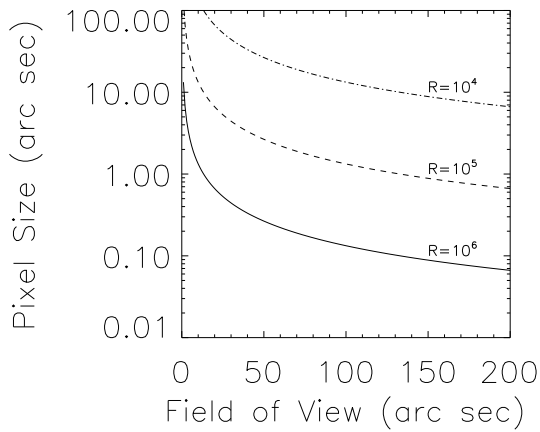,width=3.4in,angle=90}}
\caption{\small Pixel size as a function of the field of view required to
spatially fully sample fringes at the edge of the FPA, and hence
maintain modulation efficiency.  Curves are plotted for resolutions $R
= k/\delta k = 10^4$, $10^5$ and $10^6$ assuming an 8~m diameter
primary aperture and a beam splitter of diameter 10~cm.
}
\label{pixel_fov}
\end{figure}

\noindent
Fig. \ref{pixel_fov} shows the pixel size for a given field of view
for a range of resolutions.  For example, for an 8~m diameter primary
aperture and a beam splitter of diameter 10~cm, a FOV of $200''$, and
a pixel size of $0.''05$, the corresponding angles at a beam splitter
of diameter 10~cm are $2.^\circ 2$ and $4''$ respectively, leading to
a resolution limit of $R = 1.3 \times 10^6$.  Since this resolution is
two orders of magnitude greater than we are proposing, it is clear
that spectral resolution is not the principal factor determining pixel
size.  An alternative way to view this constraint is that $d$, i.e.,
the size of the optics, is determined not by spectral resolution, but
by the requirement that there be no vignetting over the field of
view. Thus, the optics for an IFTS are similar to that of a simple
re-imaging camera, and are smaller and slower than those of an
equivalent dispersive spectrograph.  

We therefore have broad freedom to choose the pixel size by trading
off field of view and spatial sampling.  Given that NIR arrays of
$4096 \times 4096 $ pixels are likely to be available in the near
future, a pixel size of $0.''05$ yields a $3.'3$ field of view and
$\lambda / 2D$ sampling at 4 $\mu$m.  This choice of pixel size does
not preclude diffraction limited imaging at shorter wavelengths. If
pixels have sharp boundaries, then it is possible to extract
information at spatial frequencies above the cut-off in the
pixel-sampling modulation transfer function if the spacecraft can
offset and track at the sub-pixel level (cf. Fruchter and Hook 1998).
Similar reasoning suggests $0.''1$ pixels would be a satisfactory
compromise for the MIR channel.

\section{SNR CALCULATIONS}
\label{snrcal}

An IFTS views all frequencies in its pass band simultaneously, but
multiplexes them by modulating each optical frequency in the source at
an ``acoustic'' frequency proportional to the optical frequency. Near
the zero phase difference point of the interferometer, assuming 
an ideal beamsplitter, full intensity
is transmitted through one port of a dual port interferometer, and no
intensity through the other. For reasonably high resolution, most of
the points in the interferogram are acquired away from the
centerburst, and thus, to a good approximation, the average intensity
transmitted through {\em each port} is 50\% of the source intensity.
In the time domain, the average signal photoelectron count rate for a
dual-port interferometer is

\begin{equation}
\dot{N_s}  = \int_{k_{min}}^{k_{max}}
\eta (k)  S_k dk 
\end{equation}

\noindent
where $S_k$ is the source photon rate per unit wavenumber, $\eta$ is
the system efficiency, including the telescope, collimator,
beam-splitter, and camera throughput and the detector QE. The integral
is taken over the full bandpass of the system. The signal-to-noise
ratio in the time domain is given, on average, by

\begin{eqnarray}
SNR_t & = & {\dot{N_s}t \over  
\left[  t \int_{k_{min}}^{k_{max}}   
\eta(k) (S_k + B_k)dk  + 2(tI_d  + n_r^2)   \right]^{1/2}} \\
\end{eqnarray}

\noindent
Where the noise consists of photon shot noise from source and
background photon rate, $B_k$, dark current $I_d$, and read noise,
$n_r$. The integration time per OPD step is $t$. The factor of two in dark
and read noise occurs because of the twin FPAs required for a
dual-port instrument.

The relation between the signal to noise ratio in the spectral domain
to the signal to noise ratio in the temporal domain can most easily be
derived using Parseval's theorem.  If there are $N$ frames in the
interferogram the relation between the noise level in the spectral
domain, $\sigma_k$, to the noise in the time domain, $\sigma_t$,
assuming that it is approximately white is given by

\begin{equation}
|\sigma_k  |^2 = N^{-1} |\sigma_t|^2.
\end{equation}

\noindent
Hence, the SNR in the frequency domain is 

\begin{equation}
SNR_k = N^{1/2} SNR_t/M ,
\end{equation}

\noindent
where the last equality is obtained under the assumption of a white
spectrum extending over $M$ spectral channels.
The SNR for an IFTS
at any given resolution simply scales as $1/M$ per spectral
channel.

For all SNR calculations (Table \ref{capabilities} and numbers in \S
\ref{science}) we have assumed a dual-port design on a 8-m telescope
and $\eta = 0.5$. The background consists of zodiacal light at 1 AU
and thermal emission from the telescope.  The internal background
within the IFTS is orders of magnitude lower than the external
background because we expect that the IFTS optics and detectors will
operate in the same low temperature ($\simeq$ 30~K) environment within
the NGST instrument bay.  Therefore the internal background has been
neglected.  We have used the emission measured at the ecliptic pole
(Hauser et al. 1984), and include thermal background from the
telescope optics, assuming $T=50$~K and an emissivity of $0.06$.  The
SNR results simulate optimal extraction of synthetic aperture
photometry of a point source from a digital image. The size of the
aperture which maximizes the SNR depends on the dominant noise source and
the wavelength, and ranges between $\theta_{50} = 1.07 \lambda/D$, i.e.,
the angular diameter which encircles 50\% of the light,
and
$\theta_{80} = 1.79 \lambda/D$.

The instrument performance depends on the detector dark current and
read noise.  For the NIR channel we assume a dark current of 0.03
e~s$^{-1}$ and an rms read noise of 5 e$^-$ [with Fowler sampling,
(Fowler \& Gatley 1990)]. For the MIR channel we assume a dark current
of 1 e$^-$~s$^{-1}$ and an rms read noise of 5 e$^-$. This performance
is optimistic, but not unrealistic given projected detector
development for NGST (Bely \& McCreight 1996).

\section{NGST-IFTS SCIENCE}
\label{science}

The versatility, broad wavelength coverage, and spatial multiplexing
capability of an IFTS renders it well suited to executing a large
space telescope's broad range of science goals. In this section, we
discuss the applications of an IFTS which are representative of the
many programs that can be carried out with this instrument.

\subsection{Galaxy Formation}

\begin{figure}[thb]
\centerline{\psfig{figure=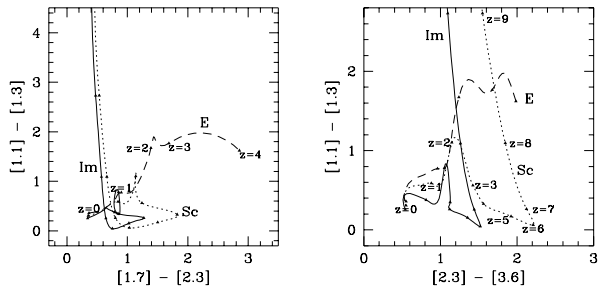,width=3.4in,angle=0}}
\vskip -1.8 in
\caption{\small Colors for Im, Sc and E galaxies as a function of redshift in
five 1600~cm$^{-1}$ wide pass-bands centered at 1.1, 1.3, 1.7, 2.3, \&
3.6 $\mu$m.  Colors are plotted from $z$=0-10, a triangle is plotted
at every interval of unit redshift.  These curves assume unevolving
models and only account for the K-corrections and the intergalactic
absorption due to HI (this is only important at $z>8$ at 1.1 $\mu$m).
Thus the low resolution IFTS colors provide a powerful measure of the
age of stellar populations and photometric redshifts.  In a typical
$10^5$~s exposure ($K = 29.5$, SNR=10) the IFTS can do a good job of
separating high-z objects from foreground objects.  }
\label{color_color}
\end{figure}

Since the rest-frame optical emission of distant galaxies is
redshifted to the NIR, broad-band, wide-field IR imaging surveys are
essential to the study of their formation, early evolution, and
merging history.  The NIR colors synthesized from a low resolution 5
spectral channel IFTS survey offer excellent separation of galaxy type
and redshift throughout the range $z=1$ to $z<10$
(Fig. \ref{color_color}).  Thus, high-$z$ galaxies can be picked out
from foreground objects, and a preliminary determination of their
stellar populations made. An IFTS provides this capability along with
the flexibility to conduct much higher spectral resolution surveys.

\begin{figure}[thb]
\centerline{\psfig{figure=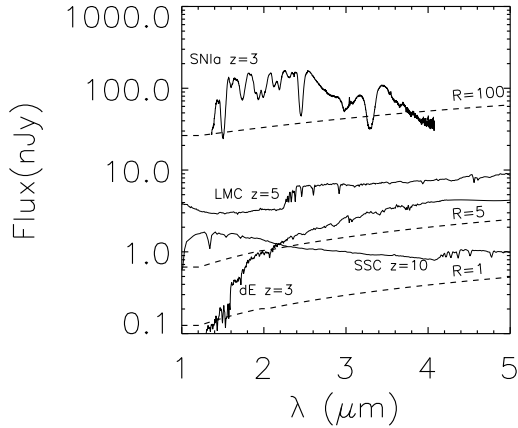,width=3.4in,angle=90}}
\caption{\small Sensitivity of IFTS in a $10^5$~s exposure.  See \S
\ref{snrcal} for details.  The full 1-5 $\mu$m band-pass is scanned,
and the flux corresponding to SNR=10 is plotted as a dashed line. 
Spectral resolutions
of $M=5$ spectral channels (broad-band imaging) and $M=100$ spectral
channels (low resolution spectroscopy) are shown. Also shown are the
spectra of a type Ia supernova (SNIa), 
a LMC-like Magellanic irregular ($M_B = -18.7$), a
dwarf elliptical ($M_B = -15$) and a super star-cluster (SSC),
representative of a proto-globular cluster, forming stars for 25 Myr
at 1 $M_\odot$~yr$^{-1}$ $H_0 = 50$~km~s$^{-1}$~Mpc$^{-1}$, and $q_0 =
0.1$. }
\label{survey1}
\end{figure}

With an integration time of $10^5$ s per $3.'3$ FOV, an IFTS will
obtain a SNR=5-15 at the 1 nJy flux level in each of the 5 spectral
channels, and measure dwarf (rest-frame $M_B = -18.7$) star forming
irregular galaxies at $z=5$ at a SNR=40, and detect LMC-like
star-forming systems or super star-clusters which might be
representative of proto-globular clusters forming stars at 1
$M_\odot$~yr$^{-1}$ for 25 Myr out to $z\approx 10$, and, if they
exist, dwarf ellipticals ($M_B = -15$) to $z=3$ (Fig. \ref{survey1}).

Extrapolating the $K$-band galaxy luminosity function (see
Fig. \ref{number-counts}), taking only passive evolution into account,
predicts that a $10^5$~s $M=5$ IFTS observation will yield high SNR
($\ge 6$) multi-color photometry for about 11,000 objects per field
(for $q_0$=0.1; for $q_0$=0.5, there should be at least 3500 sources).
Hence, in this mode, an IFTS can probe the evolution of the luminosity
function as a function of morphological type and stellar content, and
thereby determine the spectral and merging history of galaxies. In
addition, collapsing the interferogram yields an exquisitely deep (20
pJy rms!) 1--5 $\mu$m broad-band image, enabling a morphological study
of the faintest and very lowest surface brightness sources.

In a low-resolution spectroscopic mode (e.g., $M=100$ spectral
channels), an IFTS can yield spectroscopic redshifts (accurate to 
$\delta z \simeq 0.02$),
explore the stellar population age and measure gas abundances in
galaxies, and make extinction-free measurements of the star-formation
history of the universe. For a typical exposure of $10^5$ s, an $M=100$
IFTS observation reaches $K=25.7$ (35 nJy) at SNR=10 per spectral
channel, and should detect about 4500 sources to this limit in a
single FOV ($q_0 = 0.1$). 
The spectral resolution is sufficient to detect standard
HII region diagnostics (e.g., Pa$\alpha$ at $z$=0--2, H$\alpha$ \&
[OIII] at $z$=0.5--10; Ly$\alpha$ at $z>6$) and well-studied stellar
features (e.g., 4000\AA, 2900\AA\ and 2640\AA\ spectral breaks). At
the detection limit (40 nJy), IFTS will be sensitive to star
formation rates as low as 1--10 ${\rm M_\odot\ yr^{-1}}$ for galaxies
at $z$=3--7 (Kennicutt 1983). Moreover, H$\alpha$ is a more robust
measure of the star-formation rate than are rest-frame UV
diagnostics, since it is relatively insensitive to dust
extinction. For example, the intrinsic star-formation rates in the
$z$=2.5--3.5 Lyman drop galaxies are likely to be $>100{\rm M_\odot\
yr^{-1}}$ --- factors of a few above the rates estimated from their
rest-frame 1500\AA\ luminosity (Steidel et al.~1996). As such, the
$M=100$ IFTS survey will provide an accurate measure of the
star-formation history of the Universe to $z\simeq 5$. For $L^*$
($M_B=-21$) star-forming galaxies, the 2640\AA, 2900\AA\ and 4000\AA\
spectral breaks can be measured to SNR=40 out to $z=6$, yielding a
measure of the mean (luminosity-weighted) age of the stellar
population with an accuracy of $<0.5$~Gyr. Population synthesis
modeling of the observed spectra can yield even more accurate measures
of the relative ages and stellar content of galaxy samples, and
potentially provide an estimate for the first epoch of galaxy
formation and a lower limit to the age of the universe.

\subsection{Evolution of IR Galaxies}

While only one third of the bolometric luminosity of local galaxies is
radiated in the IR (Soifer \& Neugebauer 1991), there is growing
evidence that this fraction is actually increasing with redshift.  The
deepest counts available from IRAS at 60 $\mu$m (Hacking \& Houck 1987),
which correspond to an average redshift of about 0.2 (Ashby et
al. 1996), already suggest some evolution of the IR emission in the
universe.  A deep survey with the ISO at 15 $\mu$m has discovered a
few objects at $z$=0.5-1 with star formation rates much higher than
deduced from the optical (Rowan-Robinson et al. 1997). These
conclusions are reinforced by unexpectedly high far-IR and sub-mm
source counts measured by ISO and the JMCT/SCUBA (Puget et al. 1997;
Smail et al. 1997).

Deep optical surveys (Lilly et al. 1995; Williams et al. 1996) probe
the rest frame UV luminosities of high redshift galaxies, which can be
converted into star formation rates under plausible assumptions about
young stellar populations. Analysis of these data suggests that the
star formation rate of the universe peaked at $z\simeq 1$ and then
declined (Lilly et al. 1996; Madau 1996).  This has led to claims that
the primary epoch of star formation in the universe has been
seen. However, the conversion of UV luminosities into star formation
rates must take into account a correction for the luminosity fraction
absorbed by the dust which is generally associated with young stars.
Since this correction is uncertain for high redshift galaxies, the
star formation rates currently deduced from optical surveys alone
might be substantially underestimated (Calzetti 1997).  
For high redshift galaxies,
the only current direct observational constraint is set by the recent
detection of the cosmic far-IR background built up from the
accumulated IR light of faint galaxies along the line of sight. The
far-IR and sub-mm background light detected by COBE implies a star
formation rate which is a factor of two above that inferred from
optical galaxies in the Hubble Deep Field (Puget et al. 1996; Schlegel
et al. 1997).

\begin{figure}[thb]
\centerline{\psfig{figure=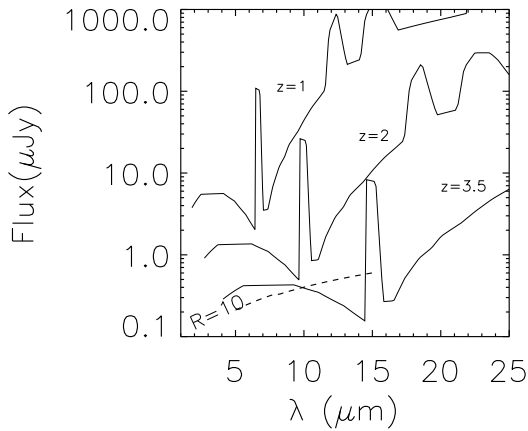,width=3.4in,angle=90}}
\caption{\small MIR spectra of an ultraluminous ($10^{12}~L_\odot$) IR galaxy
are plotted for $z=1-3.5$ (Guiderdoni et al. 1997), representative of
objects such as Arp 220.  The most prominent narrow feature is the 3.3
$\mu$m PAH emission band.  The sensitivity of IFTS in a $10^4$~s
exposure is shown.  The full 5-15 $\mu$m band-pass is scanned, and the
flux corresponding to SNR=10 for $M=10$ spectral channels is plotted
as a dashed line.  $H_0 = 50$~km~s$^{-1}$~Mpc$^{-1}$, and $q_0 =
0.1$.}
\label{midir-plot}
\end{figure}

Mid-IR low resolution spectroscopy can be used to search for
dusty, star forming galaxies at high redshift.
Fig. \ref{midir-plot} shows a sequence of redshifted spectra
representing an ultraluminous ($10^{12}~L_\odot$) IRAS galaxy such as
Arp~220, which is a prototype for a deeply embedded starburst
(Guiderdoni et al. 1997).  (Arp~220 is not an AGN, as recent ISO
spectroscopy and VLBI observations show (Genzel et al. 1997; Smith,
Lonsdale, \& Lonsdale 1998).)  The most prominent feature in these
spectra is 3.3 $\mu$m PAH emission which is shifted to the MIR band.
Low resolution MIR spectral scans will be exquisitely sensitive to this
broad emission feature, and therefore provide a sensitive way to
search for deeply extincted star-formation at high-$z$.  For example,
SIRTF will survey 1 square degree in the far-IR to a
flux level of about 100 $\mu$Jy to study the cosmological evolution of
these sources. This detection threshold is sufficient to find about
200 ultraluminous IR galaxies to $z\simeq 3$. A NGST follow-up
of the SIRTF square degree survey, lasting 1 month (330 fields
$\times$ $10^4$~s), would detect all these galaxies at high SNR in
addition to many objects with lower star formation rates. The
detection of redshifted PAH emission provides a unique signature of
embedded star formation, and protects against confusion with
galactic cirrus, that limits the usefulness of longer wavelength
searches.

\subsection{Large-scale Structure}

The study of galaxy clustering is a classical test of theories of
cosmic structure formation and a means of discriminating among
cosmological world models.  Since an IFTS acquires simultaneously
both spatial and redshift information, it is the ideal instrument for
the exploration of 3-dimensional clustering.  The $3.'3$ FOV of our IFTS
is well-matched to the size of clusters of galaxies at $z>1$.

Foreground contamination will be a serious obstacle to the
identification of high-$z$ clusters, and obtaining spectra of very
many (faint) galaxies is critical to establishing cluster membership
and richness. Without redshifts, clusters might easily go unnoticed.
Since an IFTS obtains a spectrum for every pixel in the FPA, it is
suited to the discovery and study of clusters.  In particular,
obtaining spectra of galaxies with complex morphologies (where slit
placement would be difficult or wasteful of light) is straightforward
with an IFTS.  Examples include young galaxies that appear as a
collection of small proto- globular cluster-sized clumps undergoing
bursts of star formation (e.g., Tegmark et al. 1997), or galaxies
undergoing merging.  

IFTS spectral scans with 1000 spectral channels are well-suited to
probing the velocity dispersion of rich clusters ($\wig> 1000$~
km~s$^{-1}$).  In a $10^5$~s scan, an IFTS can obtain spectra throughout
the 1-5 $\mu$m region of objects at the 800 nJy level (SNR=5), i.e.,
$L^*$ star-forming galaxies at $z\simeq2$ or $L^*$ ellipticals, if
they exist, at $z\simeq3$.  Emission line diagnostics such as [OIII]
and H$\beta$ are well-suited to probing the cluster velocity field. In
addition, absorption lines such as NaI can be used to study the
interstellar absorption of cluster galaxies (cf. Steidel et al. 1996)

Since we will simultaneously obtain high SNR rest-frame optical/UV
morphology for these objects, it will be possible to study the
virialization process and merging history of high-$z$ clusters.  These
results, while interesting in and of themselves, will also complement
future X-ray studies of the cluster virialization process to be
carried out with AXAF.

Since a spectrum is obtained for every pixel, IFTS studies of
clusters will also detect and obtain spectra for 
gravitational lens arcs and images.
While the morphology and surface brightness structure of lensed images
can be used to reconstruct the cluster mass distribution, their
spectra can be used to probe the properties of very distant galaxies
at high SNR (e.g., Franx et al. 1997).

\subsection{Star and Planet Formation}

Simultaneous broad wavelength studies of star forming regions enable
the study of stars, brown dwarfs, and planets at short wavelengths ($
\lambda < 5~\mu$m) and their formation environments at longer
wavelengths ($ \lambda > 5~\mu$m).  NGST can observe low-mass stars in
star-forming regions out to several kpc, and will 
map out the detailed properties of protoplanetary disks
as a function of age, stellar mass, and environment in many
star-forming regions with a total sample of thousands of stars.  The
unique ability of an IFTS to obtain colors and spectroscopy over a
wide wavelength range for every object in the field makes it a
powerful tool.  Colors, reddening, luminosity and spectral
classification for every object will discriminate against cluster
non-members, allow construction of Hertzsprung-Russell diagrams, and
provide determination of cluster ages, age spreads, and measurement of
the initial mass function.

The combination of the NIR and MIR channels of an IFTS can be used to
determine the frequencies and lifetimes of protoplanetary disks and to
understand the evolution of their dust and gas.  While NIR excess
emission traces hot dust near (0.1 AU) the star, the MIR is critical
for probing material at planet formation distances.  Spectra from
1-15~$\mu$m at $M=50$ will provide detailed spectral energy
distributions and yield radial disk structure, reveal gaps due to the
presence of protoplanets, and determine dust composition from
solid-state emission features.  For stars in which active accretion
has ceased, and in older clusters, an IFTS will provide sensitive MIR
measurements of optically thin dust disks, the precursors of $\beta$
Pic-like systems.  Spectroscopy of the MIR rotational lines of H$_2$
will be used to determine the relative gas and dust dispersal
time-scales and place limits on the time for the formation of giant
gas planets.

Star formation regions are ideal for the study of the sub-stellar mass
function and isolated super-planets due to the high stellar density
and the brightness of sub-stellar objects in their youth.  An exciting
prospect is the study of extra-solar giant planets (EGPs) that have
been ejected from young planetary systems.  The unexpected discovery
that some Jupiter-mass planets orbiting nearby stars have highly
eccentric orbits (Marcy \& Butler 1996; Cochran et al. 1997) suggests
that ejection of planets by dynamical scattering is a common outcome
of the planet formation process (Lin \& Ida 1997).  These planets can
be distinguished from free-floating EGPs that formed in isolation (via
gravitational collapse) by their high proper motions that will far
exceed the velocity dispersions in young clusters.  Very young (1 Myr)
clusters are optimum for searching for ejected EGPs since they would
be luminous and not have traveled far from where they formed.  In
addition, the high stellar density ($10^2$/FOV) provides a high
probability of discovering ejected EGPs.  Planets formed at AU
distances will be ejected with velocities $\simeq$ 30 km s$^{-1}$; in
Orion this is a proper motion of 0.06" in 5 yr, easily measured during
the NGST lifetime.  The discovery of high proper-motion EGPs will
provide an unique opportunity to study, via $M\simeq$ 100 spectroscopy,
the atmospheres of {\it true Jupiter analogs} (i.e., planets with a
formation history similar to that of our own solar system) without the
usual difficulties of studying orbiting planets in the glare of the
central star.

An IFTS on NGST can be used to measure the mass function for
sub-stellar objects from brown dwarfs to super-planet masses on
account of its ability to obtain photometry and spectroscopy of
distant ($>$1 kpc), young open clusters.  Such clusters are well
matched to the FOV of our IFTS, and $M\simeq100$ spectroscopy will
provide spectral classification via H$_2$O and CH$_4$ bands of several
hundred very low-mass stars and sub-stellar objects per cluster. At an
age of 10 Myr, an IFTS can study free-floating super-planets of 5
$M_{Jup}$ at a distance of 1 kpc in $10^5$~s.  Comparison of the
atmospheric compositions implied by the spectral properties of ejected
EGPs and those formed in isolation may confirm their different
formation histories.

\subsection{Kuiper Belt Objects}

Kuiper Belt objects (KBOs) hold great significance for our
understanding of the formation and evolution of the solar system, both
as a source of short period comets and as primitive remnants of the
planet-building phase of solar system history.  One of the primary
barriers to the detailed study of KBOs is that ground-based surveys do
not reach deep enough limiting magnitudes to accumulate significant
samples of objects.  In addition, the high proper motion of KBOs
(0.5-2.5"/hour for objects at 40 AU) requires that accurate orbital
parameters must first be derived for precise slit placement in
spectroscopic follow-up observations.

An IFTS offers the sensitivity and multi-object capability that is
perfectly suited to discovery and follow up spectroscopy of numerous,
faint, high proper motion targets.  In an IFTS surveys of the ecliptic
plane, KBOs would be detected by their characteristic proper motion
observed in a series of images taken at each successive OPD.  The
corresponding interferograms for each detected object may be
appropriately re-registered and stacked for Fourier transform recovery
of the spectrum for each source.  The spectrum of KBOs is a clue to
their surface composition and collisional history.

Assuming a typical $R-K$ color for KBOs (e.g., Tegler \& Romanishin
1997) and extrapolating the number counts at $R=23$ (Jewitt, Luu, \&
Chen 1996) to the nJy level, an IFTS scan of $10^5$~s over 5 spectral
channels between 1-5 $\mu$m is sufficient to detect about 40 KBOs per
$3.'3$ FOV at SNR=10.  Much higher spectral resolution scans are
feasible for brighter objects.  From these data we would be able to
construct a detailed census of objects between the orbit of Neptune
and the Oort cloud, and thereby provide a direct observation of the
solar nebula and unique constraints on the dynamical history of the
solar system.

\section{CONCLUSIONS}

An IFTS instrument can perform a wide variety of NGST science.  The
advantages of the IFTS concept are:

\begin{enumerate}

\item Deep imaging acquired simultaneously with higher spectral
resolution data over a broad wavelength range.

\item ``Hands-off'', unbiased, multi-object, slitless spectroscopy
(ideal for moving objects).  Efficient in confusion limit.

\item Flexible resolution ($M=1-10,000$).

\item High throughput (near 100\%) dual-port design.

\item Tolerant of cosmic rays, read-noise, dark current, and light leaks.

\item Simple and reliable calibration.  High SNR determination of
flat-fields and detector non-linearity.

\item Compact, lightweight design. Slow reimaging optics.

%\item Proven technology with low development costs.

\end{enumerate}

\acknowledgments

%Thanks to XXX.

\end{document}